\documentclass[11pt]{article}
\usepackage{cmap}
\usepackage[utf8]{inputenc}
\usepackage[T2A]{fontenc}
\usepackage[english]{babel}
\inputencoding{utf8}
\usepackage{amsmath,amssymb,amsthm,graphicx}
\usepackage{authblk}
\usepackage{xcolor}
\usepackage[pdfpagemode=UseOutlines,
bookmarks=true, bookmarksopenlevel=3, bookmarksopen=true,
bookmarksnumbered=true, unicode=true,
pdffitwindow=true,pdfpagelayout=SinglePage,pdfhighlight=/P,
colorlinks=true,linkcolor=teal,citecolor=blue,urlcolor=blue]{hyperref}

\advance\textwidth 60mm  \advance\oddsidemargin -30mm
\advance\textheight 47mm \advance\topmargin -35mm

\theoremstyle{definition}
\newtheorem{definition}{Definition}
\theoremstyle{remark}
\newtheorem*{remark}{Remark}

\theoremstyle{plain}
\newtheorem{prop}{Proposition}

\title{The negative symmetry classification problem}

\author[]{M.P. Kolesnikov\thanks{kolesnikov.mp@phystech.edu}} 
\affil[]{Center of Integrable Systems, P.G. Demidov Yaroslavl State University, Yaroslavl, Russia}

\date{}

\begin{document}

\maketitle
\begin{abstract}
A negative symmetry is a nonlocal symmetry of special type. In this paper, we introduce a method for constructing negative symmetries from consistent triplets of differential and differential-difference equations. Moreover, we study the relation between 3D consistent equations in the discrete case and  the continuous case. 
\end{abstract}
\bigskip
\begin{quotation}
\end{quotation}

\section{Introduction}

Symmetries play an important role in the theory of integrable systems. A system is called integrable if it has an infinite number of symmetries \cite{Shabat_1991}.  By "symmetry" one usually means local symmetry, but nonlocal symmetries are also significant with plenty of applications \cite{Lou_2012,Lou_2024}. A negative symmetry is a nonlocal symmetry of a special type. In particular, negative symmetries are associated with string equations \cite{Kol,Fok,Moor}. Moreover, certain negative symmetries are known fundamental integrable equations \cite{Ad_3D_24}.  Therefore, there is a need to study negative symmetries, which may give rise to important integrable equations. 

Negative symmetries can be obtained from consistent chains as shown in Section \ref{neg_sec}. In particular, negative symmetries are differential consequences of consistent chains. For example, a negative symmetry of the Korteweg–de Vries (KdV) equation can be obtained by excluding a discrete variable from consistent chains.

This paper is concerned with the study of consistent chains, dressing chains for KdV-type equations that are symmetries for Volterra-type chains and of their connections to negative symmetries. Some cases of consistent equations have been investigated in the literature \cite{Garif_end,Habib_2021,Yam_2006}.  However, the connection of triples of differential and differential-difference equations with negative symmetries is not described in the literature. 

We formalize the ideas presented in the article \cite{Ad_3D_24} and the results of this paper are the following:
\begin{itemize}
    \item We give definitions of the consistency of equation pairs and conditions which are equivalent to the consistency of equation pairs.
    \item We show how to use these conditions to obtain new equations and
    conditions for the dressing chains of KdV-type equations.
    \item We show how to obtain negative symmetries from consistent chains.
    \item We explain the relation between the 3D consistency of equations in the discrete case and continuous case.
\end{itemize}

This paper is organized as follows.

In the next section, we introduce the notation we use throughout the text, and we give the definitions of negative symmetries and explain the relations between them and recursion operators. 

In Section \ref{Compatible equations}, we give definitions of consistency of studied equations and obtain conditions that are equivalent to the consistency of these equations. These conditions are used to make a general proposition about dressing chains of KdV-type equations. Moreover, we present a general scheme for constructing negative symmetries from consistent equations.  

In section \ref{Examples}, we present examples of consistent triples of equations, we check the consistency condition in examples, and obtain negative symmetries for them.

Section \ref{3D compatibility} deals with the relation between consistency of B\"acklund transformations and 3D compatible equations in the discrete case. The second part of this section shows how 3D compatibility in the continuous case is connected with the discrete case.

Finally, in Section \ref{Conclusion}, we summarize the results and present some ideas for future work.

\section{Preliminaries}
In this section, we describe negative symmetries and their relation with recursion operators and consistent equations. 

Throughout the text:
\begin{itemize}
  \item By $x,z,t,\tau$ we denote continuous variables, by $u,v,w$  functions of independent continuous and discrete variables $n\in\mathbb{Z},\:u=u(n)=u_n$.  By $u_{x},u_{xx},u_{xxx}$ we denote derivatives $\partial_xu,\partial^2_xu,\partial^3_xu.$
  \item Functions of dependent and independent variables will be denoted by $P,Q,F,R,G,f,g,h,p,q$,  constants and parameters will be denoted by $a,b,c,\alpha,\beta...$
  \item By $\mathcal{R}$ we denote a recursion operator. By $D,D^{-1}$ we denote a total derivative operator and the inverse operator.  By $T$ we denote a shift operator, applying $T$ to a function $f(u_n,\partial_xu_n\ldots)$,  we obtain $T(f(u_n,\partial^k_xu_n,\ldots))=f(T(u_{n}),T(\partial_x^ku_{n}),\ldots)=f(u_{n+1},\partial_x^ku_{n+1},\ldots).$ 
\end{itemize}
An equation
 $$u_\tau=f(x,u,u_x,\ldots, \partial^k_xu)$$ 
 is called a symmetry for an equation 
 $$u_t=g(x,u,u_x,\ldots, \partial^m_xu)$$ if it satisfies the following relation 
 $$(u_t)_\tau-(u_\tau)_t=(g(x,u,u_x,\ldots, \partial^m_xu))_\tau-(f(x,u,u_x,\ldots, \partial^k_xu))_t=0.$$
A recursion operator is an operator that transforms symmetries into new symmetries. It plays an important role in mathematical physics, allowing one to construct infinite sequences of symmetries and to prove that some systems are integrable in the symmetry sense. 

It is known that the operator
\[
\mathcal{R}=D^2-4u-2u_xD^{-1}
\]
is a recursion operator for the KdV equation $u_t=u_{xxx}-6uu_x$.

Consider a function (we denote $u_z$) satisfies the following equation
\begin{equation}\label{neg_KdV}
    \mathcal{R}(u_z)=(D^2-4u-2u_xD^{-1})(u_z)=-4\alpha u_z 
\end{equation}
it is equivalent to the following system 
\begin{equation}\label{neg_KdV_2}
    u_z=q_x,\quad q_{xxx}-4(u-\alpha)q_x-2u_x q=0
\end{equation}

Motivated by this example, we give the following definition. 
\begin{definition}Consider an integrable system with one dependent function $u$ and a recursion operator $\mathcal{R}$. An equation
\begin{equation}
    u_z=\tilde{G}(x,u,\partial_xu,\ldots,\partial^n_xu,q^1,\partial_xq^1,\ldots,q^2,\ldots ,q^3,\ldots   )
\end{equation}
is called a negative symmetry if it satisfies the following equation 
\begin{equation}\label{Rec}
    \mathcal{R}(u_z)=\alpha u_z, 
\end{equation}
where $(q^1,q^2,\ldots)$ are nonlocal variables, which may satisfy additional equations 
$$A_i(x,u,\partial_xu,\ldots,\partial^n_xu,q^1,\partial_xq^1,\ldots, q^2,\ldots,q^3,\ldots)=0.$$
\end{definition}
Negative symmetries are often related to well-known equations of mathematical physics \cite{Aratyn}. A negative symmetry of mKdV is related to the sine-Gordon equation \cite{Ad_neg_flow_2024} , a negative symmetry of KdV to the Camassa-Holm equation \cite{Hone_1999,Schiff_1998}, for the nonlinear Schrödinger equation it is related to the Maxwell–Bloch system \cite{Pavlov_2002}. Moreover, negative symmetries are related to non-autonomous reductions of the Painleve type \cite{Kol,Ad_Vol,Orlov}. Negative symmetries are also generating functions for higher symmetries of the equation \cite{Lou_2012,Lou_2024}.  

System (\ref{neg_KdV_2}) is a negative symmetry for KdV. After integration of  the second equation with the factor $2q$ and the substitution  $2v_x=u,2v_z=q$, we obtain the following equation:
\begin{equation}\label{neg_mkdv}
    v_{xxz}=\frac{v_{xz}^{2}-\beta^{2}}{2v_{z}}+2\left(2v_{x}-\alpha\right)v_{z}.
\end{equation}
Motivated by this example, we give the following definition. Consider equations of the form
\begin{equation}\label{KdV_type}
     u_{t}=u_{xxx}+f(x,u,u_x,u_{xx})=F\left(x,u,u_x,u_{xx},u_{xxx}\right),
\end{equation}
\begin{equation}\label{neg}
u_{xxz}=G\left(x,u,u_{x},u_{xx},u_{z},u_{zx}\right).
\end{equation}
\begin{definition} Equation (\ref{neg})  is called a negative symmetry for an  equation of KdV-type (\ref{KdV_type}) if it satisfies the following relation: 
\begin{equation}
    \left(D_{x}^{2}D_{z}-D_{G}\right)\left(u_{xxx}+f\left(x,\ldots\right)\right)=B\left(u_{xxz}-G\left(x,\ldots\right)\right),
\end{equation}
\end{definition}
where $D_G$ is the Fréchet derivative and B is a differential operator $B=a_0D^3+a_1D^2+a_2D+a_3$ and $a_{i}$ are some functions of $x,u,u_{x},u_{xx}\ldots$. 

In certain cases, equation (\ref{neg}) can be integrated to yield a hyperbolic equation of the form $u_{xz}=a\left(u,u_{x},u_{z}\right)$. Such equations, which admit KdV-type evolutionary symmetries, may be regarded as the simplest class of negative symmetries. An exhaustive classification of these equations was obtained in the paper \cite{Sok_2011}, but the results of Sokolov show that hyperbolic equations alone are insufficient. In particular, neither the KdV equation itself nor several other evolution equations are compatible with any hyperbolic equations. Furthermore, hyperbolic equations do not describe the general solutions of equation (\ref{Rec}), but rather special ones—typically corresponding to the case where the parameter $\alpha$ vanishes. The negative symmetry in Definition 2 may not be equivalent to that in Definition 1, and vice versa, for instance, the negative symmetry generated by the sixth-order recursion operator \cite{Demskoi_2008} for the Krichever-Novikov equation does not reduce to the form $u_{xxz}$.

One can show that we obtain an equation of the form (\ref{neg}) by eliminating the discrete variable from the equations 
\begin{equation}\label{dress_h}
    h(x,u_n,u_{n,x},u_{n+1},u_{n+1,x})=0,
\end{equation}
\begin{equation}\label{Vol_pr}
    u_{n,z}=g(u_n,u_{n+1},u_{n-1}).
\end{equation}
where we assume $\partial_{u_{n+1}}h\neq0,\partial_{u_{n-1}}h\neq0$. Such chains were examined in the papers \cite{Habib_2015,Habib_2014,Garif_end}.

Now, one can show that there are KdV-type equations (\ref{KdV_type}), equations (\ref{dress_h})(dressing chian) and equations (\ref{Vol_pr})(Volterra-type chain), that are consistent with each other.  And by eliminating the discrete variable we obtain a negative symmetry for the KdV-type equation \cite{Ad_3D_24}. Finally, we have the following equations:
\begin{equation}\label{list}
    \begin{array}{lcc}
     u_{t}=u_{xxx}+f(x,u,u_x,u_{xx})=F\left(x,u,u_x,u_{xx},u_{xxx}\right),  \\
   \\
    u_{n+1,x}=\tilde{h}_+\left(x,u_{n},u_{n,x},u_{n+1},\alpha\right),\\
    \\
u_{n,z}=g\left(u_{n-1},u_{n},u_{n+1}\right),\\
   \\
    u_{xxz}=G(x,u,u_x,u_{xx},u_z,u_{xz},\alpha),\\
    \\
\end{array}
\end{equation}
Here the dressing chain contains a parameter $\alpha$, which corresponds to the parameter in equation (\ref{Rec}).
\section{Compatible equations}\label{Compatible equations}
In this section, we give a definition of consistency of the equations from the  list (\ref{list}) and present a method for constructing such equations.
\subsection{Compatibility of KdV and dressing chain}
Consider a third order evolution equation  
$$  u_{t}=F\left(x,u,u_x,u_{xx},u_{xxx}\right),$$
and a differential difference equation of hyperbolic type
 \[
    u_{n+1,x}=h_+\left(x,u_{n},u_{n,x},u_{n+1}\right).
    \]
In this case, convenient variables are the following $\left(v,u,u_1,u_2,\ldots\right)$, where we denote $v= u_{n+1},u= u_{n},u_1=u_{n,x},u_{2}=u_{n,xx},\ldots$.

Consider the following vector fields which generate flows defined by the corresponding equations 
\begin{equation}\label{D_x}
    D_{x}=D=\partial_{x}+h_+\left(x,v,u,u_{x}\right)\partial_{v}+u_{1}\partial_{0}+u_{2}\partial_{1}+\ldots
\end{equation}
\begin{equation}\label{D_t_1}
    D_{t}=\partial_{t}+T\left(F\right)\partial_{v}+F\partial_{0}+D\left(F\right)\partial_{1}+\ldots,
\end{equation}
where $\partial_i=\partial_{u_i}$ and $T$ is the shift operator $T\left(u_k\right)=D^k(v)$. Taking into account the above notations, we give the following definition: 
\begin{definition}
    Equations 
    \begin{equation}\label{KdV_def}
         u_{t}=F\left(x,u,u_1,u_{2},u_{3}\right),
    \end{equation}
    \begin{equation}\label{dress_def}
        v_x=h_+\left(x,v,u,u_1\right)
    \end{equation}
 are called consistent  if the following equations hold identically: 
    \begin{equation}\label{def_1}
    \left[T,D_{x}\right]=0, \quad  \left[D_t,D_{x}\right]=0,\quad  \left[T,D_{t}\right]=0.
    \end{equation}
In (\ref{def_1}) the first equation is fulfilled on functions of the form $g(u,u_1,u_2\ldots)$, equations (\ref{def_1}) are equivalent to the following equation:
 \begin{equation}\label{con_1}
    TD_x\left(F\right)=D_t\left(h_+\right).
    \end{equation}
\end{definition}
For two given functions (\ref{KdV_def},\ref{dress_def}), one can check condition (\ref{con_1}). For unknown functions one can check that equation (\ref{con_1}) is linear in $u_4$ and the coefficient at $u_4$ is equal to    
\begin{equation}\label{coef_1}
    T\left(\partial_{u_3} F\right)-\partial_{u_3} F.
\end{equation}
Condition (\ref{con_1}) holds identically. Therefore, the coefficient is equal to zero. For KdV-type equations, it is indeed equal to zero. Moreover, we will assume that the dressing chain has no pseudo constants. That is, only constants are solutions to the following equations $T(C)=C,\:D(C)=0$. Using this, we obtain that equation(\ref{KdV_def}) is linear in the third derivative. Such equations have been studied in detail, and there is a complete list of integrable KdV-type equations \cite{Sok_2014}
\[
\begin{array}{l}
u_t = u_{xxx} + uu_x, \\
\\
u_t = u_{xxx} + u^2u_x, \\
\\
u_t = u_{xxx} + u^2_x, \\
\\
u_t = u_{xxx} - \frac{1}{2}u_x^3 + (c_1e^{2u} + c_2e^{-2u})u_x, \\
\\
u_t = u_{xxx} - \frac{3u_xu_{xx}^2}{2(u_x^2+1)} + d_1(u_x^2+1)^{3/2} + d_2u_x^3, \\
\\

u_t = u_{xxx} - \frac{3u_{xx}^2}{2u_x} + \frac{1}{u_x} - \frac{3}{2}\wp(u)u_x^3, \\
\\

u_t = u_{xxx} - \frac{3u_xu_{xx}^2}{2(u_x^2+1)} - \frac{3}{2}\wp(u)u_x(u_x^2+1), \\
\\

u_t = u_{xxx} - \frac{3u_{xx}^2}{2u_x}, \\
\\
u_t = u_{xxx} - \frac{3u_{xx}^2}{4u_x} + c_1 u_x^{3/2} + c_2 u_x^2, \quad c_1 \neq 0 \text{ or } c_2 \neq 0, \\
\\
u_t = u_{xxx} - \frac{3u_{xx}^2}{4u_x} + \alpha(x)u_x, \\
\\
u_t = u_{xxx} - \frac{3u_{xx}^2}{4u_x} + \frac{3}{\xi} u_{xx}(\sqrt{\alpha'u_x} + u_x) + \frac{3u_x^3}{\xi^2} + \frac{6}{\xi^2} u_x^{5/2}\sqrt{\alpha'} \\
\quad + \frac{3u_x^{3/2}}{\xi^2\sqrt{\alpha'}} (\xi\alpha'' - 2\alpha'^2) + s u_x + c_0 + c_1 u + c_2 u^2, \\
\text{where } \xi = \alpha(x) - u, s = -\frac{\alpha'''}{\alpha'} + \frac{3\alpha''^2}{4\alpha'^2} + \frac{3\alpha''}{\xi} - \frac{\alpha'^2}{\xi^2} - \frac{c_0 + c_1 \alpha + c_2 \alpha^2}{\alpha'}, \\
\\
u_t = u_{xxx} + 3u^2 u_{xx} + 9uu_x^2 + 3u^4 u_x + u_x \alpha(x) + \frac{1}{2} u \alpha'(x), \\
\\
u_t = u_{xxx} + 3uu_{xx} + 3u_x^2 + 3u^2 u_x + (u \gamma(x))_x + \beta(x), \\
\\
u_t = u_{xxx} + \alpha(x)u_x + \beta(x)u.
\end{array}
\]
Here $(\wp')^2 = 4\wp^3 - g_2\wp - g_3$, where $a_1, a_2, c_0, c_1, c_2, g_2, g_3$ are arbitrary constants, and $\alpha, \beta$, and $\gamma$ are arbitrary functions.

Considering differential-difference equations with pseudo-constants allows us to obtain new examples of negative symmetries and consider evolutionary equations that are not linear in the highest derivative.

There are equations in the list that are of the following form
\begin{equation}\label{Kdv_prop_1}
    u_{t}=u_{3}+f\left(x,u,u_{1}\right).
\end{equation}
One can show that  the coefficient of $u_3$ in the consistency condition (\ref{con_1}) is 
\[
\partial_{u_{1}}\partial_xh_++h_+\partial_{v}\partial_{u_{1}}h_++u_{1}\partial_{u}\partial_{u_{1}}h_++u_{2}\partial_{u_{1}}\partial_{u_{1}}h_+=D\left(\partial_{u_{1}}h_+\right)=0.
\]
Therefore, we obtain the following. 
\begin{prop}\label{prop_1}
    For equations  (\ref{Kdv_prop_1}) any consistent dressing chain has the following form:
    \[
    h_+(x,v,u,u_1)=a\:u_1+h_2(x,v,u).
    \]
\end{prop}
 Moreover, one can prove a more general proposition.
\begin{prop} Consider the following equations compatible in sense of (\ref{con_1}) 
\[
u_{t}=u_{3}+f_1\left(x,u,u_{1}\right)u_{2}^{2}+f_2\left(x,u,u_{1}\right)u_{2}+f_3\left(x,u,u_{1}\right),
\]
\[
v_1=h_+(x,v,u,u_1).
\]
 Then $v_1$ satisfies the following equations, for different $f_1(x,u,u_1)$    
\[
\begin{array}{llr}
f_1  =  0,& v_1 = h_3u_1 + h_4, & \text{(A)} \\
f_1 =  -\dfrac{3}{4u_1},& v_1^{1/2} = h_3u_1^{1/2} + h_4, \; h_4 \neq 0, & \text{(B)} \\
f_1  =  -\dfrac{3}{2u_1},& u_1v_1 = h_3, & \text{(C)} \\
f_1=  -\dfrac{3u_1}{2(u_1^2 + s(x, u))}, \; s \neq 0, & sv_1^2 - 2h_3u_1v_1 + T(s)u_1^2 = h_3^2 - sT(s),& \text{(D)}
\end{array}
\]
 where $h_3=h_3(x,u,v),\:h_4=h_4(x,u,v)$ and $s=s(x,u)$ are some functions. 
\end{prop}
Consistent equations (\ref{dress_def},\ref{KdV_def}) have been studied in the paper \cite{Garif_end}, where dressing chains have been derived that are consistent with KdV-type equations that are integrable in the symmetry sense.

\subsection{Dressing chains and Volterra-type chains }

Consider the differential difference equation of hyperbolic type
\begin{equation}\label{dress_int}
    u_{n+1,x}=h_+\left(x,u_{n},u_{n,x},u_{n+1}\right)
\end{equation}
and the Volterra-type chain
\begin{equation}\label{vol}
u_{n,z}=g\left(u_{n-1},u_{n},u_{n+1}\right).
\end{equation}
Consider these equations in some arbitrary $m$. In this case, the dynamic variables are the following $\left(w,u,u_{\pm1},u_{\pm2},\ldots\right)$, where we denote $u_m=u,u_{m\pm k}=u_k,u_{m,x}=w,\:\partial_{u_{m\pm k}}=\partial_{\pm k}$. Other variables and their derivatives can be expressed in terms of the dynamic variables. 
Moreover, assume that equation (\ref{dress_int}) can be written as
\begin{equation}\label{dress_minus}
    u_{-1,x}=h_-\left(x,u_{},w,u_{-1}\right).
\end{equation}
Then the shift operator is defined as follows:
$$T^{k}\left(u_{n}\right)=u_{n+k},\:T\left(w\right)=h_+\left(x,u,u_1,w\right),\:T^{-1}\left(w\right)=h_-\left(x,u,u_{-1},w\right)\quad k\in \mathbb{Z}.$$
Consider vector fields which generate flows defined by the corresponding equations 
\begin{equation}
    D=D_{x}=w\partial_{0}+T\left(w\right)\partial_{1}+T^{-1}\left(w\right)\partial_{-1}+T^{2}\left(w\right)\partial_{2}+\ldots,
\end{equation}
\begin{equation}
    D_{z}=D\left(g\right)\partial_{w}+g\partial_{0}+T\left(g\right)\partial_{1}+T^{-1}\left(g\right)\partial_{-1}+T^{2}\left(g\right)\partial_{2}+\ldots.
\end{equation}
Taking into account the above notations, we give the following definition: 
\begin{definition}
Equations:
\[
u_z=g(u,u_1,u_{-1}),
\]
\[
u_{1,x}=h_+(x,w,u,u_1),
\]
 are called consistent  if the following equations hold identically:
\[
 \left[D_{z},D_{x}\right]=0,\quad \left[D_{x},T\right]=0,\quad \left[D_{z},T\right]=0
\] 
which are equivalent to the following equation:
\begin{equation}\label{condition_2}
    DT\left(g\right)=D_{z}\left(h_+\right).
\end{equation}
\end{definition}
For two given equations (\ref{dress_minus},\ref{vol}), one can check condition (\ref{condition_2}). 

The integrability condition for discrete equations was introduced in the paper \cite{Levi_2009}. Volterra-type chains have been studied in detail and there is the complete list of integrable Volterra-type chains \cite{Yam_2006}
\begin{align*}
u_z &= P(u)(u_1 - u_{-1}),&  & \text{(V1)} \\
\\
u_z &=  P(u^2) \left( \frac{1}{u_1 + u} - \frac{1}{u + u_{-1}} \right), & & \text{(V2)} \\
\\
u_z&  =  G(u) \left( \frac{1}{u_1 - u} + \frac{1}{u - u_{-1}} \right),&  & \text{(V3)} \\
\\
u_z & =  \frac{R(u_1, u, u_{-1}) + v R(u_1, u, u_1)^{1/2} R(u_{-1}, u, u_{-1})^{1/2}}{u_1 - u_{-1}},&  & \text{(V4)} \\
\\
u_z & =  y(u_1 - u) + y(u - u_{-1}), & y' = P(y)\quad & \text{(V5)} \\
\\
u_z & =  y(u_1 - u)y(u - u_{-1}) + \mu, & y' = P(y)/y\quad & \text{(V6)} \\
\\
u_z & =  (y(u_1 - u) + y(u - u_{-1}))^{-1} + \mu, & y' = P(y^2)\quad & \text{(V7)} \\
\\
u_z & =  (y(u_1 + u) - y(u + u_{-1}))^{-1}, & y' = G(y) \quad & \text{(V8)} \\
\\
   u_z & =  \frac{y(u_1 + u) - y(u + u_{-1})}{y(u_1 + u) + y(u + u_{-1})}, & y' = P(y^2)/y\quad & \text{(V9)} \\
   \\
u_z & =  \frac{y(u_1 + u) + y(u + u_{-1})}{y(u_1 + u) - y(u + u_{-1})}, & y' = G(y)/y\quad & \text{(V10)} \\
\\
u_z & =  \frac{(1 - y(u_1 - u))(1 - y(u - u_{-1}))}{y(u_1 - u) + y(u - u_{-1})} + \mu, & y' = \frac{P(y^2)}{1 - y^2}\quad & \text{(V11)} 
\end{align*}
where $\nu\in\left\{0,\pm1 \right\}$, the functions $P(u),G(u),R(u,v,w)$ are polynomials of the form 
\[
P(u)=\alpha u^2+\beta u+\gamma,
\]
\[
G(u)=\alpha u^4+\beta u^3+\gamma u^2+\eta u+\zeta, 
\]
\[
R(u,v,w)=(\alpha v^2 + 2\beta v + \gamma) uw + (\beta v^2 + \lambda v + \delta)(u + w) + \gamma v^2 + 2\delta v + \varepsilon.
\]
The coefficients of $P, G, R$ and the variable $\mu$ are arbitrary constants.

Consistent equations (\ref{dress_minus},\ref{vol}) have been studied in the paper \cite{Garif_end}, where  Volterra-type chains have been derived, which are consistent with dressing chains of certain KdV-type equations.

\subsection{Compatibility of KdV and Volterra}
Consider a third order evolution equation  
\begin{equation}\label{KdV_3}
    u_{n,t}=F\left(x,u_n,u_{n,x},u_{n,xx},u_{n,xxx}\right),
\end{equation}
and a Volterra-type chain
\begin{equation}\label{vol_3}
u_{n,z}=g\left(u_{n-1},u_{n},u_{n+1}\right).
\end{equation}
In this case, equations (\ref{KdV_3},\ref{vol_3}) are evolutionary equations, and the dynamical variables are all variables \\ $(u_n,u_{n,x},u_{n,xx},\ldots ,u_{n\pm1}, u_{n\pm1,x},\ldots)$ 
Consider vector fields which generate flows defined by the corresponding equations 
\begin{equation}\label{D_z}
    D_{z}=D=g\partial_{u_n}+D_x(g)\partial_{u_{n.x}}+\ldots +T^{\pm1}(g)\partial_{u_{n\pm 1}}+\ldots
\end{equation}
\begin{equation}\label{D_t}
    D_{t}=F\partial_{u_n}+D_x(F)\partial_{u_{n.x}}+\ldots +T^{\pm1}(F)\partial_{u_{n\pm 1}}+\ldots
\end{equation}
\begin{definition}
    Equations (\ref{KdV_3},\ref{vol_3}) are called consistent  if the following equation holds identically: 
    \begin{equation}\label{def_3}
    \left[D_z,D_t\right]=0, 
    \end{equation}
which are equivalent to the following equation:
 \begin{equation}\label{con_3}
    D_z\left(F\right)=D_t\left(g\right).
    \end{equation}
\end{definition}
For two given functions (\ref{KdV_3},\ref{vol_3}), we can check condition (\ref{con_3}). 
\begin{equation}\label{condition_3}
    \left(u_{n,xxx}+f\left(x,u_{n},u_{n,x},u_{n,xx}\right)\right)_{z}=\left(g\left(u_{n-1},u_{n},u_{n+1}\right)\right)_{t}.
\end{equation}
it is equal to an expression of the form $$A\left(x,u_n,u_{n+1},u_{n-1},u_{n,x},u_{n+1,x}\ldots u_{n-1,xxx}\right)=0,$$
where the dots denote the remaining derivatives in $x$ up to the third order.  The last equation cannot be satisfied via equations (\ref{KdV_3},\ref{vol_3}).  However, if we additionally have a dressing chain (\ref{dress_int}) consistent with equations (\ref{KdV_3}) and (\ref{vol_3}), we can substitute the derivatives, starting from the highest-order ones.  Equation (\ref{condition_3}) holds if equations (\ref{KdV_3},\ref{vol_3}) are consistent via the dressing chain.

\subsection{Negative symmetry}\label{neg_sec}
Consider consistent equations, a differential difference equation of hyperbolic type
\begin{equation}\label{dress_neg}
     \tilde{h}(x,u_n,u_{n,x},u_{n+1},u_{n+1,x},\alpha)=0,
\end{equation}
and a Volterra-type chain
\begin{equation}\label{vol_neg}
u_{n,z}=g\left(u_{n-1},u_{n},u_{n+1}\right).
\end{equation}
We assume that equation (\ref{dress_neg}) can be uniquely solved for variables $u_{n+1,x},u_{n,x}$ and equation (\ref{vol_neg}) for variables $u_{n-1},u_{n+1}.$

We differentiate the Volterra-type chain (\ref{vol_neg}) with respect to $x$:
\begin{equation}
    u_{n,zx}=g_{1}\left(u_{n},u_{n,x},u_{n+1},u_{n-1},u_{n+1,x},u_{n-1,x}\right).
\end{equation}
Now, we substitute  the derivatives of $u_{n\pm1}$ given by the dressing chain (\ref{dress_neg}) and $u_{n-1}$ by equation (\ref{vol}) and we obtain the following equation
\begin{equation}\label{uzx}
    u_{n,zx}=g_2(u_n,u_{n,x},u_{n,z},u_{n+1},\alpha).
\end{equation}
One can write equation (\ref{uzx}) as
\begin{equation}\label{un1}
    u_{n+1}=g_3(u_n,u_{n,x},u_{n,z},u_{n,zx},\alpha).
\end{equation}
Finally, we substitute equation (\ref{un1}) to the dressing chain (\ref{dress_neg}) and obtain the following equation
\begin{equation}\label{neg_neg}
    u_{n,xxz}=G(x,u_n,u_{n,x},u_{n,xx},u_{n,z},u_{n,xz},\alpha).
\end{equation}
Moreover, if equations (\ref{dress_neg},\ref{vol_neg}) are consistent with a KdV-type equation, then equation (\ref{neg_neg}) is a negative symmetry. However, in certain cases where consistent Volterra-type and hyperbolic-type chains lack continuous symmetries, equation (\ref{neg_neg}) remains suitable for consideration.
\section{Examples}\label{Examples}
In this section, we demonstrate the definitions and methods using examples
from the paper \cite{Ad_3D_24}.

Consider the potential Korteweg–de Vries equation (pot-KdV)
\begin{equation}\label{pKdV}
    u_{t}=u_{xxx}-6u^2_{x}. 
\end{equation}
It is known that pot-KdV is consistent with the following dressing chain
\begin{equation}\label{dress_pot}
    u_{n+1,x}+u_{n,x}=\left(u_{n+1}-u_{n}\right)^{2}+\alpha\:.
\end{equation}
One can check it by using condition (\ref{con_1})
\begin{equation}\label{con_pot}
    TD\left(u_{xxx}-6u^2_{x}\right)=D_{t}\left(\left(v-u\right)^{2}-u_x+\alpha\right).
\end{equation}
One can check that the right-hand side of condition (\ref{con_pot}) is equal to the following $8\alpha^2u_0+8u_0^3u_1-16u_0u_1^2-8v^3(\alpha+u_1)+8\alpha(u_0^3-u_0u_1)+8v^2(3\alpha u_0+3u_0u_1-u_2)-8u_0^2u_2+12u_1u_2+4u_0u_3-4v(2\alpha^2+6\alpha u_0-2\alpha u_1+6u_0^2u_1-4u_1^2-4u_0u_2+u_3)-u_4 $, and the left-hand side is equal to the same. Therefore, condition (\ref{con_pot}) holds identically.

Morevoer, it is known that the dressing chain (\ref{dress_pot}) is consistent with the following Volterra-type chain
$$u_{n,z}=\frac{\beta}{u_{n+1}-u_{n-1}}. \eqno(V_4^*)$$
One can check it by using condition (\ref{condition_2})
 \begin{equation}\label{con_pot_2}
     DT\left(\frac{\beta}{u_{1}-u_{-1}}\right)=D_{z}\left(\left(u_1-u\right)^{2}-v+\alpha\right).
 \end{equation}
The left-hand side of condition (\ref{con_pot_2}) is equal to $D\frac{\beta}{u_{2}-u}=\frac{\beta v}{(u_{2}-u)^{2}}-\frac{\beta\left(-(u_{1}-u)^{2}+(u_{2}-u_{1})^{2}+v\right)}{(u_{2}-u)^{2}}$ and the right-hand side of condition (\ref{con_pot_2})  is equal to $D_{z}(a)=-\frac{2\beta(u_{1}-u)}{u_{1}-u_{-1}}+\frac{2\beta(u_{1}-u)}{u_{2}-u}-\frac{\beta\left(\alpha+(u-u_{-1})^{2}-v\right)}{(u_{1}-u_{-1})^{2}}+\frac{\beta\left(\alpha+(u_{1}-u)^{2}-v\right)}{(u_{1}-u_{-1})^{2}}$ bringing to a common denominator we obtain that the left-hand side is equal to the right-hand side. Therefore, condition (\ref{con_pot_2}) holds identically.

Following the scheme from section \ref{neg_sec} we solve the dressing chain with respect to $u_{n-1,x}$
$$u_{n-1,x}=\left(u_{n}-u_{n-1}\right)^{2}-u_{n,x}+\alpha.$$
Differentiate the Volterra-type chain 
$$u_{nx,z}=\frac{\beta  (u_{n-1}-2 u_n+u_{n+1})}{u_{n-1}-u_{n+1}},$$
where we substituted the derivatives of $u$ and simplify. From the last equation and the Volterra-type chain we obtain  
$$u_{n+1}=u_n-\frac{u_{n,\text{xz}}-\beta }{2 u_{n,z}}.$$
Finally, we substitute $u_{n+1}$ into the dressing chain and obtain the following  negative symmetry 
$$u_{xxz}=\frac{u_{xz}^{2}-\beta^{2}}{2u_{z}}+2\left(2u_{x}-\alpha\right)u_{z}. $$

Consider the potential modified Korteweg–de Vries equation (pot-mKdV)
\begin{equation}\label{pMKdV}
    u_{t}=u_{xxx}-2u_{x}^{3}.
\end{equation}
It is known that pot-mKdV is consistent with the following chains
$$u_{n+1,x}+u_{n,x}=\alpha\cosh\left(u_{n+1}-u_n\right),\quad u_{z}=c\frac{e^{u_{n+1}}+e^{u_{n-1}}}{e^{u_{n+1}}-e^{u_{n-1}}}-\beta.$$
One can construct the following negative symmetry by excluding the discrete variable
$$u_{xxz}=2u_{x}\sqrt{u_{xz}^{2}+\alpha^{2}\left(u_{z}+\beta\right)^{2}-\alpha^{2}c^{2}}-\alpha^{2}\left(u_{z}+\beta\right).$$

Finally, consider the Krichever-Novikov Equation (the KN equation) 
\begin{equation}\label{KN}
    u_{t}=u_{xxx}-\frac{3\left(u_{xx}^{2}-q\left(u\right)\right)}{2u_{x}},\quad q(u)=c_{4}u^{4}+c_{3}u^{3}+c_{2}u^{2}+c_{1}u+c_{0}.
\end{equation}
Consider a symmetric biquadratic polynomial $h(u,v)$.  $\partial^3_u h(u,v)=\partial_v^3h(u,v)=h(u,v)-h(v,u)=0$. Then the following chains 
\begin{equation}\label{KN_chain}
    u_{n,x}u_{n+1,x}=h\left(u_{n+1},u_{n}\right),\quad u_{n,z}=\frac{2h\left(u_{n},u_{n+1}\right)}{u_{n+1}-u_{n-1}}-h^{\left(0,1\right)}\left(u_{n},u_{n+1}\right)
\end{equation}
are consistent  with each other and with the KN equation (\ref{KN}), here  $q(u)=(\partial_vh\left(u,v\right))^{2}-2h\left(u,v\right)\partial_vh\left(u,v\right)$.

One can construct the  following  negative symmetry by excluding the discrete variable from (\ref{KN_chain}):
\begin{multline}\label{neg_KN}
P(u)(u_x u_{xxz} - u_{xx}u_{xz})^2 - u_x^2(P'(u)u_{xz} - \\
-(4u^2 - 8\alpha u - 8\alpha^2 + g_2)u_x u_z)(u_x u_{xxz} - u_{xx} u_{xz}) + \\
+(2\beta u_x^2 - P(u))(r(u)u_{xz}^2 - r'(u)u_x u_z u_{xz} + 4(2u + \alpha)u_x^2 u_z^2) + \\
+4u_x^2((u - \alpha)u_{xz} - u_x u_z)^2 - 16\gamma^2 u_x^2(\beta u_x^2 - P(u))^2 = 0,
\end{multline}
where $P(u)=u^4+\frac{1}{2}g_2u^2+2g_3u+\frac{1}{16}g_2^2-\alpha q(u)$. There are other ways to write down the negative symmetry (\ref{neg_KN}) that are convenient in different applications \cite{Ad_3D_24}.

Another important example is the Burgers equation
\begin{equation}\label{Burger}
    u_{t}=u_{xx}+2uu_{x},
\end{equation}
which is related to the equations 
\begin{equation}\label{Pot_Burger}
    v_{t}=v_{xx}+v_{x}^{2}
\end{equation}
\begin{equation}\label{heat}
    w_{t}=w_{xx}
\end{equation}
via the following substitutions
\begin{equation}\label{subs_b}
    u=v_{x}=\frac{w_{x}}{w},\quad v=\log w
\end{equation}
The heat equation (\ref{heat}) admits a recursion operator given by $\mathcal{R}_{w}=D_{x}=D$. Recursion operators for equations (\ref{Pot_Burger},\ref{Burger}) can be derived via $\mathcal{R}_{u}=D\mathcal{R}_{v}D^{-1},\mathcal{R}_{v}=w^{-1}\mathcal{R}_{w}w$.
\[
\mathcal{R}_{u}=D+u+u_{x}D^{-1},\quad\mathcal{R}_{v}=D+v_{x}
\]
Using these recursion operators, we derive negative symmetries 
\begin{equation}
    \begin{array}{c}
w_{xz}=\alpha w_{z}+\beta w_{x}+\gamma w,\\
v_{xz}=-v_{x}v_{z}+\alpha v_{z}+\beta v_{x}+\gamma,\\
u_{xz}=\frac{u_{x}\left(u_{z}-\alpha\beta-\gamma\right)}{u-\alpha}-\left(u-\alpha\right)u_{z}.
\end{array}
\end{equation}
Moreover, the negative symmetry for the heat equation arises from the following chains
\[
w_{n,x}+a_{1}w_{n+1,x}=a_{2}w_{n}+a_{3}w_{n+1},\quad w_{n,z}=b_{1}w_{n+1}+b_{2}w_{n}+b_{3}w_{n-1}
\]
where $a_{1}a_{3}=a_{2}$.

Furthermore, by applying substitutions (\ref{subs_b}), we can derive analogous chains for the Burgers potential version and B\"acklund transformation for Burgers equation.
\begin{equation}
    \begin{array}{c}
e^{v_{n}}v_{n,x}+a_{1}e^{v_{n+1}}v_{n+1,x}=a_{2}e^{v_{n}}+a_{3}e^{v_{n+1}},\qquad v_{n,z}=b_{1}e^{v_{n+1}-v_{n}}+b_{2}+b_{3}e^{v_{n-1}-v_{n}},\\
\frac{u_{n+1,x}}{u_{n+1}-a_{3}a_{1}^{-1}}-\frac{u_{n,x}}{u_{n}-a_{2}}+u_{n+1}-u_{n}=0.
\end{array}
\end{equation}
Since these negative symmetries were derived via recursion operators, they automatically serve as negative symmetries for the higher-order symmetries of equations (\ref{Burger},\ref{Pot_Burger},\ref{heat}).
\begin{equation}
    \begin{array}{c}
u_{t_{3}}=u_{xxx}+3uu_{xx}+3u_{x}^{2}+3u^{2}u_{x},\\
v_{t_{3}}=v_{xxx}+3v_{x}v_{xx}+v_{x}^{3},\\
w_{t_{3}}=w_{xxx}.
\end{array}
\end{equation}

\section{3D compatibility}\label{3D compatibility}
In this section,  we show how  to obtain 3D consistent equations from dressing chains and give a definition of the 3D consistency of equations in the continuous case.
\subsection{Discrete case }
Consider the dressing chain for pot-KdV
\begin{equation}\label{dress_3D}
    \partial _xu_{n+1}+\partial _xu_{n}=\left(u_{n+1}-u_{n}\right)^{2}+\alpha\:.
\end{equation}
Equation (\ref{dress_3D}) is a Bäcklund transformation. Therefore, we can consider the commutativity of the two transformations with different parameters. We assume that the transformations with different parameters act on different discrete variables 
\begin{equation}\label{dress_3D_1}
    \partial _xu_{n+1,m}+\partial_x u_{n,m}=\left(u_{n+1,m}-u_{n,m}\right)^{2}+\alpha_1,
\end{equation}
\begin{equation}\label{dress_3D_2}
   \partial _xu_{n,m+1}+\partial_x u_{n,m}=\left(u_{n,m+1}-u_{n,m}\right)^{2}+\alpha_2.
\end{equation}
Now, consider the following transformations:
\begin{equation}\label{dress_3D_3}
   \partial _xu_{n+1,m+1}+\partial_x u_{n,m+1}=\left(u_{n+1,m+1}-u_{n,m+1}\right)^{2}+\alpha_1,
\end{equation}
\begin{equation}\label{dress_3D_4}
   \partial _xu_{n+1,m+1}+\partial_x u_{n+1,m}=\left(u_{n+1,m+1}-u_{n+1,m}\right)^{2}+\alpha_2.
\end{equation}
One can check that the system of equations (\ref{dress_3D_1},\ref{dress_3D_2},\ref{dress_3D_3},\ref{dress_3D_4}) is compatible if the following equation holds
\begin{equation}\label{H_1}
    \left(u_{n,m}-u_{n+1,m+1}\right)\left(u_{n+1,m}-u_{n,m+1}\right)=\alpha_{1}-\alpha_{2}.
\end{equation} 
Equation (\ref{H_1}) connects values of the function $u$ in  the vertices of square.  

One can show that, if we consider equation (\ref{H_1}) in a three-dimensional lattice, the values of the function $u$ in the cube vertices do not depend on the order of computation \cite{ABS}. Therefore, equation (\ref{H_1}) is 3D compatible. Moreover, every 3D compatible equation is compatible in any dimension. Equation (\ref{H_1}) is the equation $H_1$ from the Adler-Bobenko-Suris(ABS) list \cite{ABS}. The equations from this list are multi-affine quad-equations that are 3D compatible.

Now consider the dressing chains of pot-mKdV with different parameters. Then one can check that these Bäcklund transformations are compatible if the following equation holds
\begin{equation}\label{H_3}
    \alpha_{1}\left(q_{n,m}q_{n+1,m}+q_{n,m+1}q_{n+1,m+1}\right)=\alpha_{2}\left(q_{n,m}q_{n,m+1}+q_{n+1,m}q_{n+1,m+1}\right).
\end{equation}
where $q_{n,m}=\exp{u_{n.m}}$. Equation (\ref{H_3}) is the equation $H_3$ from the ABS list with the parameter $\delta$ equal to zero. Therefore, it is 3D compatible.

It is known that the dressing chain (\ref{KN_chain}) of the Krichever-Novikov equation \cite{Ad_1998} is a symmetry for the following equation 
\begin{multline}\label{Q_4}
     c_0 u_{n,m} u_{n,m+1}  u_{n+1,m+1}  u_{n+1,m} - c_1 (u_{n,m} u_{n,m+1}  u_{n+1,m+1} + u_{n,m+1}  u_{n+1,m+1}  u_{n+1,m} \\+  u_{n+1,m+1}  u_{n+1,m} u_{n,m} +  u_{n+1,m} u_{n,m} u_{n,m+1})+ c_2 (u_{n,m}  u_{n+1,m+1} + u_{n,m+1}  u_{n+1,m}) \\- c_3 (u_{n,m} u_{n,m+1} +  u_{n+1,m}  u_{n+1,m+1}) - c_4 (u_{n,m}  u_{n+1,m} + u_{n,m+1}  u_{n+1,m+1}) + \\c_5 (u_{n,m} + u_{n,m+1} +  u_{n+1,m+1} +  u_{n+1,m}) + c_6 = 0.
\end{multline}
One can check that the B\"acklund transformations of KN equation are compatible if equation (\ref{Q_4}) holds. Equation (\ref{Q_4})  is the equation $Q_4$ from the ABS list and is related to the B\"acklund transformation for the Landau–Lifshitz equation \cite{Shabat_2}.

However, the compatibility condition of B\"acklund transformations with different parameters is not always equations from the ABS list. Consider  the dressing chain for KdV
\begin{equation}
    u_{n+1,x}+u_{n,x}=(u_n-u_{n+1})\sqrt{\alpha-(u_{n+1}+u_n)}.
\end{equation}
One can check that the corresponding equation contains the square roots of functions. Therefore, it is not an equation from the ABS list. A generalization of the ABS list has been studied in the paper \cite{Atk_2014}. The symmetries of the equations from the ABC list were analyzed in \cite{Levi_2008,Tongas_2001,Xen_2011,ABS_2}.

\subsection{Continuous case }

Consider three consistent equations: a KdV-type equation, a dressing chain and a Volterra-type chain. Consider the dressing chains with different parameters $\alpha_j$.  For each parameter, there is a discrete variable $n_j$. Moreover, there are the Volterra-type chains for each $n_j$
\begin{equation}
    u_{n_j,z_j}=g(u_{n_j},u_{n_j+1},u_{n_j-1}),
\end{equation}
where $u_{n_j\pm k}=u_{n_1,n_2,\ldots ,n_j\pm k\ldots}$.  

Moreover, we have a negative symmetry for each variable $z_j$. There is the question of the compatibility of negative symmetries.  Therefore, we give the following definition. 
\begin{definition}
    Consider the following equations
\begin{equation}\label{F3D}
    u_{xxz_{i}}=G_{i}\left(u,u_{x},u_{xx},u_{xxx},u_{z_{i}},u_{xz_{i}}\right),\quad i\in I,
\end{equation}
they are called 3D compatible if there are additional equations 
\begin{equation}\label{G3d}
    u_{z_{i}z_{j}}=P_{ij}\left(u,u_{x},u_{xx},u_{z_{i}},u_{z_{j}},u_{xz_{i}},u_{xz_{j}}\right),\quad i\neq j,
\end{equation}
such that $P_{ij}=P_{ji}$ and the following equations hold identically for pairwise different $i,j,k\in I$
\begin{equation}
    \begin{array}{c}
D_{z_{i}}\left(G_{j}\right)=D_{z_{j}}\left(G_{i}\right)=D_{x}^{2}\left(P_{ij}\right),\\
D_{z_{k}}\left(P_{ij}\right)=D_{z_{j}}\left(P_{ik}\right)=D_{z_{i}}\left(P_{jk}\right)
\end{array}
\end{equation}
by virtue of the equations (\ref{F3D}) and (\ref{G3d}) and their differential consequences  $u_{xxxz_{i}}=D_{x}\left(G_{i}\right),u_{xz_{i}z_{j}}=D_{x}\left(P_{ij}\right)$.
\end{definition}
When a negative symmetry is of the form $u_{xz}=G\left(u_{xx},u_{z},u_{x},u\right)$, the definition of 3D compatibility remains unchanged, except that formula (\ref{F3D}) must be replaced by $u_{xz_{i}}=G\left(u_{xx},u_{z_{i}},u_{x},u\right)$.

Consider two examples of 3D-compatible systems from the paper \cite{Ad_3D_24}, the
negative symmetries for pot-KdV.
\begin{prop}
    Negative symmetries
   $$u_{xxz_i}=\frac{u_{xz_i}^{2}-\beta^{2}}{2u_{z}}+2\left(2u_{x}-\alpha_i\right)u_{z_i} $$
   are 3D compatible equations and additional equations are the following
   \begin{equation}\label{neg_sup}
       v_{z_{2}z_{1}}=\frac{1}{2\left(\alpha_{1}-\alpha_{2}\right)}\left(v_{z_{1}}v_{xz_{2}}-v_{z_{2}}v_{xz_{1}}\right).
   \end{equation} 
\end{prop}
\begin{remark}
    The equations (\ref{neg_sup}) is related to the Alonso-Shabat universal hydrodynamic hierarchy \cite{Alonso}. Moreover, one can check that $(v_{z_iz_j})_{z_k}=(v_{z_iz_k})_{z_j}$ holds via the equations (\ref{neg_sup}) themselves. Therefore, the equations (\ref{neg_sup})  are consistent themselves. 
\end{remark}
Now, consider the negative symmetries for pot-mKdV (\ref{pMKdV}). 
\begin{prop}
    Equations 
    $$ v_{xxz_{i}}=2v_{x}\sqrt{v_{xz_{i}}^{2}+\alpha_{i}^{2}\left(v_{z_{i}}+\beta_{i}\right)^{2}-\alpha_{i}^{2}c_{i}^{2}}-\alpha_{i}^{2}\left(v_{z_{i}}+\beta_{i}\right)$$
    are 3D compatible equations and the additional equations are the following
$$v_{z_{i}z_{j}}=\frac{2}{\alpha_{i}^{2}-\alpha_{j}^{2}}\left(v_{xz_{j}}\sqrt{v_{xz_{i}}^{2}+\alpha_{i}^{2}\left(v_{z_{i}}+\beta_{i}\right)^{2}-\alpha_{i}^{2}c_{i}^{2}}-v_{xz_{i}}\sqrt{v_{xz_{j}}^{2}+\alpha_{j}^{2}\left(v_{z_{j}}+\beta_{j}\right)^{2}-\alpha_{j}^{2}c_{j}^{2}}\right).$$
\end{prop}

The property of 3D consistency also holds for the negative symmetries of Equations (\ref{Burger},\ref{Pot_Burger}). 
\begin{prop}
    Negative symmetries
   $$v_{xz_{i}}=-v_{x}v_{z_{i}}+\alpha_{i}v_{z_{i}}+\beta_{i}v_{x}+\gamma_{i}, $$
   are 3D compatible equations and the additional equations are the following
   \begin{equation}\label{neg_sup_3}
       v_{z_{i}z_{j}}=-v_{z_{i}}v_{z_{j}}+\frac{1}{\alpha_{i}-\alpha_{j}}((\gamma_{j}+\alpha_{i}\beta_{j})v_{z_{i}}-(\gamma_{i}+\alpha_{j}\beta_{i})v_{z_{j}}+\beta_{j}\gamma_{i}-\beta_{i}\gamma_{j})
   \end{equation} 
\end{prop}

\begin{prop}
    Negative symmetries
   $$u_{xz_{i}}=\frac{u_{x}(u_{z_{i}}-\delta_{i})}{u-\alpha_{i}}-(u-\alpha_{i})u_{z_{i}} $$
   are 3D compatible equations and additional equations are the following
   \begin{equation}\label{neg_sup_4}
       u_{z_{i}z_{j}}=\left(\frac{1}{u-\alpha_{i}}+\frac{1}{u-\alpha_{j}}\right)u_{z_{i}}u_{z_{j}}+\frac{\delta_{j}(u-\alpha_{i})u_{z_{i}}}{(\alpha_{i}-\alpha_{j})(u-\alpha_{j})}+\frac{\delta_{i}(u-\alpha_{j})u_{z_{j}}}{(\alpha_{j}-\alpha_{i})(u-\alpha_{i})},
   \end{equation} 
   where $(\delta_{i}=\alpha_{i}\beta_{i}+\gamma_{i})$.
\end{prop}

\section{Conclusion}\label{Conclusion}

In this paper, we propose a method for constructing negative symmetries. The method is based on the consideration of the triples of equations. Negative symmetries (\ref{neg}) are obtained by excluding the discrete variable from the consistent chains.  

In the second part of this paper, we considered 3D compatible equations as consistency condition for the B\"acklund transformations with different parameters. Moreover, we extend the idea of 3D compatibility in the continuous case and showed that negative symmetries are 3D compatible in this sense.

The results can be extended in the following ways. One can consider the equations from the Volterra-type chains list  and obtain a consistent  differential difference equation of hyperbolic type (\ref{dress_int}) by solving condition (\ref{condition_2}). Dressing chains for KdV-type equations are known \cite{Garif_end}. One can obtain corresponding negative symmetries by excluding the discrete variable. Moreover, one can obtain Lax representations for the negative symmetries. The known negative symmetries can be used to study Painleve-type reductions (strings equations).  One can consider equations of the higher order of derivative.  One can study 3D compatible equations in the discrete case and obtain 3D compatible equations that are not from the ABS list. It is meaningful to identify negative symmetries with Darboux-integrable differential-difference chains and analyze their properties.

 \section*{Conflict of interest} The author has no conflicts to disclose.

 \section*{Acknowledgements} The work was supported by the Foundation for the Advancement of Theoretical Physics and Mathematics “BASIS”.

\end{document}